# Epistemological and bibliometric analysis of ethics and shared responsibility- health policy and IoT systems

[1]* Corresponding author: Petar Radanliev; email: petar.radanliev@oerc.ox.ac.uk; Address: Engineering Sciences Department, Oxford e-Research Centre, 7 Keble Rd, Oxford, OX1 3QG, The University of Oxford, United Kingdom, https://orcid.org/0000-0001-5629-6857

Petar Radanliev [1]*, David De Roure [1]

[1] Department of Engineering Science, Oxford e-Research Centre, 7 Keble Road, Oxford, OX1 3QG

**Conflict of Interest:** On behalf of both authors, the corresponding author states that there is no conflict of interest.

**Availability of data and materials:** all data and materials included in the article

**Funding sources:** This work was supported by the UK EPSRC [grant number: EP/S035362/1] and by the Cisco Research Centre [grant number CG1525381]. Working papers and project reports prepared for the PETRAS National Centre of Excellence and the Cisco Research Centre can be found in pre-prints online.

**Acknowledgements:** Eternal gratitude to the Fulbright Visiting Scholar Project.

**Declarations of interest:** none

**Availability of data and material:** The data used is included in the article.

**Code availability:** No code was used in this article.

**Authors' contributions:** Both authors contributed equally.

**Ethics approval:** this research has been granted ethical approval from the University of Oxford - Medical Sciences: MS IDREC Ethics Committee, under reference R64573/RE001

**Consent to participate:** Participant consent forms have been obtained from all case study participants.

**Consent for publication:** Participant information sheet was sent to all case study participants, informing the participants of the intent to publish the results of the study. The participant consent forms included a section requesting a consent for publication of anonymised and aggregated results. The submitted article has been checked (prior to submission) by all case study participants.

## Abstract

The focus in this paper is placed on shared responsibility and ethics in health policy, specific to Internet of Things (IoT) devices in healthcare systems. The article assesses how the introduction of IoT brings risks to the security of medical systems. The justification for this research emerges from the opportunities emerging from digital technologies for medical services, but also create a range of new cyber risks in the shared healthcare infrastructure. Such concerns are often not visible to individual departments in an integrated healthcare system. In addition, many healthcare organisations do not possess cyber skills and are faced with barriers imposed to adoption of smart manufacturing technologies e.g., cost. Such barriers trigger ethical concerns related to who is responsible for cyber risks in shared healthcare systems.

## 1 Introduction

This article conducts epistemological and bibliometric analysis of different topics related to shared responsibility and ethics in health policy for IoT devices and systems. Some of the present applications of IoT devices in the medical field include wearables (e.g., Apple Watch), wirelessly connected devices (e.g., like blood pressure and heart rate monitoring cuffs, glucometer), and sensors for remote monitoring and management that benefit patients, families, physicians, hospitals, and insurance companies. The advancement of IoT devices in the medical field spans into monitoring managing some of the most challenging illnesses, such as Parkinson's disease monitoring or depression and mood monitoring and includes some of the most advanced methods in the medical field, such as ingestible sensors and connected inhalers. This presents opportunities for remote monitoring and management of extremely challenging diseases and illnesses, which brings the focus of IoT devices in monitoring and managing future waves of the Covid-19 pandemic.

The aim is to use knowledge from Covid-19 to construct solutions for a digital health policy in a future Disease X event. Considering the need for periodic revision, in this study we use research records on ethics of shared responsibilities - established prior to the emergence of Covid-19 to extract knowledge, but we distinguish between the time periods with a bibliometric analysis and a case study. In the bibliometric analysis we reviewed the ethical research on shared responsibility of IoT systems. Secondly, we analysed recent studies on IoT and cyber risk. Then we used epistemological approach to relate that knowledge to ethics, shared responsibility, health policy and IoT. We advanced earlier work on ethical shared of risk responsibilities, with a new design for ethical awareness, transparency, and accountability, with a specific focus on ethics in assessing the shared risk in IoT-enabled digital healthcare systems.

## 2 Bibliometric analysis with statistical software

The statistical software used for bibliometric analysis build upon recent research on 'data mining and analysis of scientific research data records on Covid-19 mortality, immunity, and vaccine development' [1], with a specific focus on IoT, cyber risk and ethics in health policy for shared responsibilities. In other words, for the bibliometric analysis, we used the 'bibliometix' package [2], and the VOSviewer computer program [3]. Since Google scholar and other search engines were used excessively in the literature review, we searched for records on the Web of Science Core Collection. Our search for scientific literature on the topics: 'internet of things, AND ethics, AND cyber risk', produced only one record. Our second search included the same search parameters, we only replaced risk with security, this resulted with only two records. In our third search, we kept only 'internet of things, AND ethics'. This produced 88 records, which enabled us to perform bibliometric review and to seek concepts related to ethics, shared responsibility, health policy and IoT, though graphical visualisation and categorisations of data records – see Figure 1.

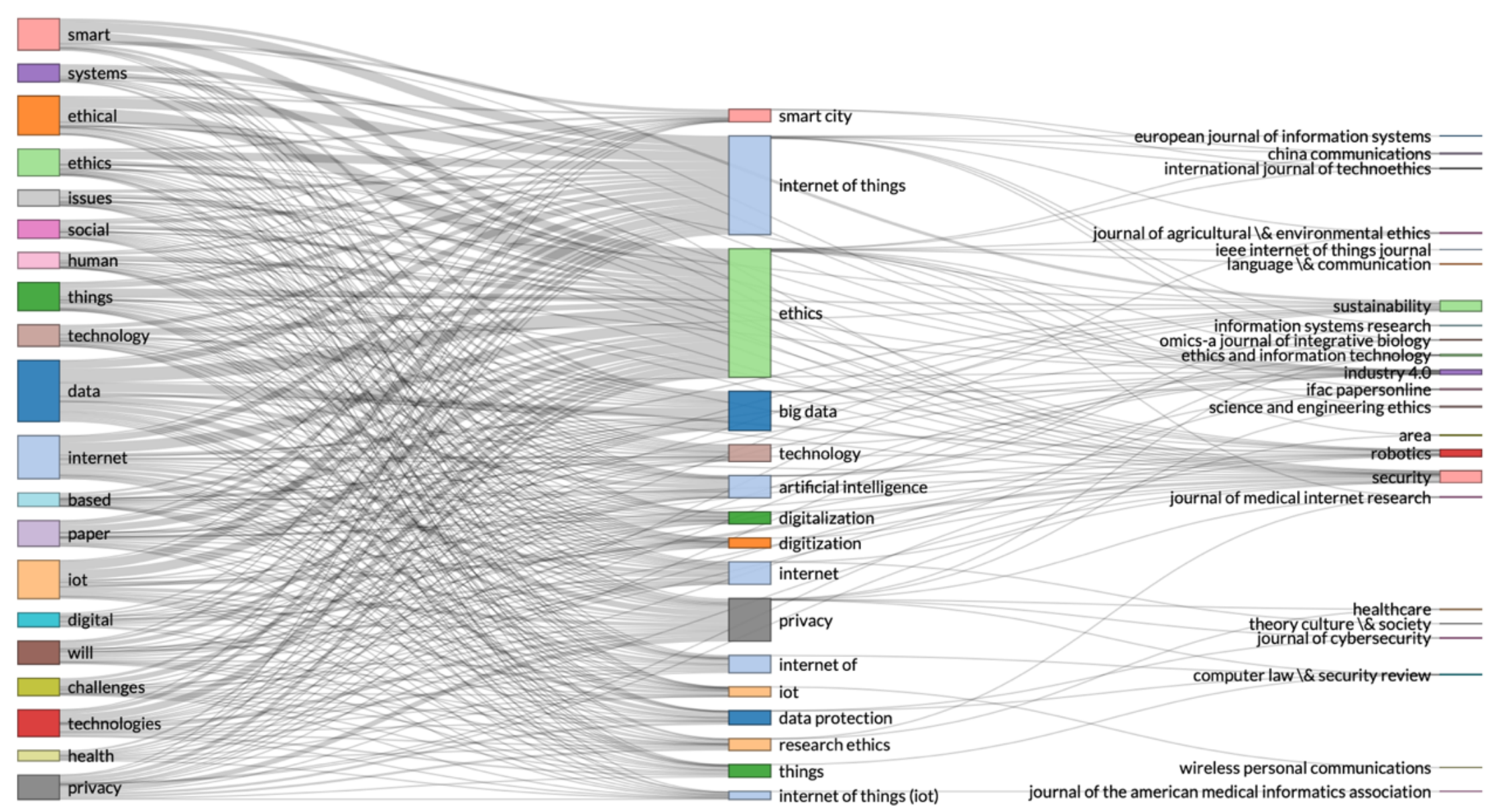

*Figure 1: bibliometric analysis - three-fields plot: keywords from abstract on the left, keywords from the text in the middle, journal sources on the right*

From the three-fields plot in Figure 1, we can see that in the data records on 'internet of things, AND ethics', the topics of shared responsibility and health policy are not represented. The only related category is healthcare. From the abstract and article keywords, extracted by the 'bibliometrix' package, the topics of privacy, data protection and risk are of relevance to health policy and IoT. In addition, the topics on ethical awareness, transparency, accountability, are missing. To analyse this data record with a different bibliometric approach, we used factorial analysis in Figure 2.

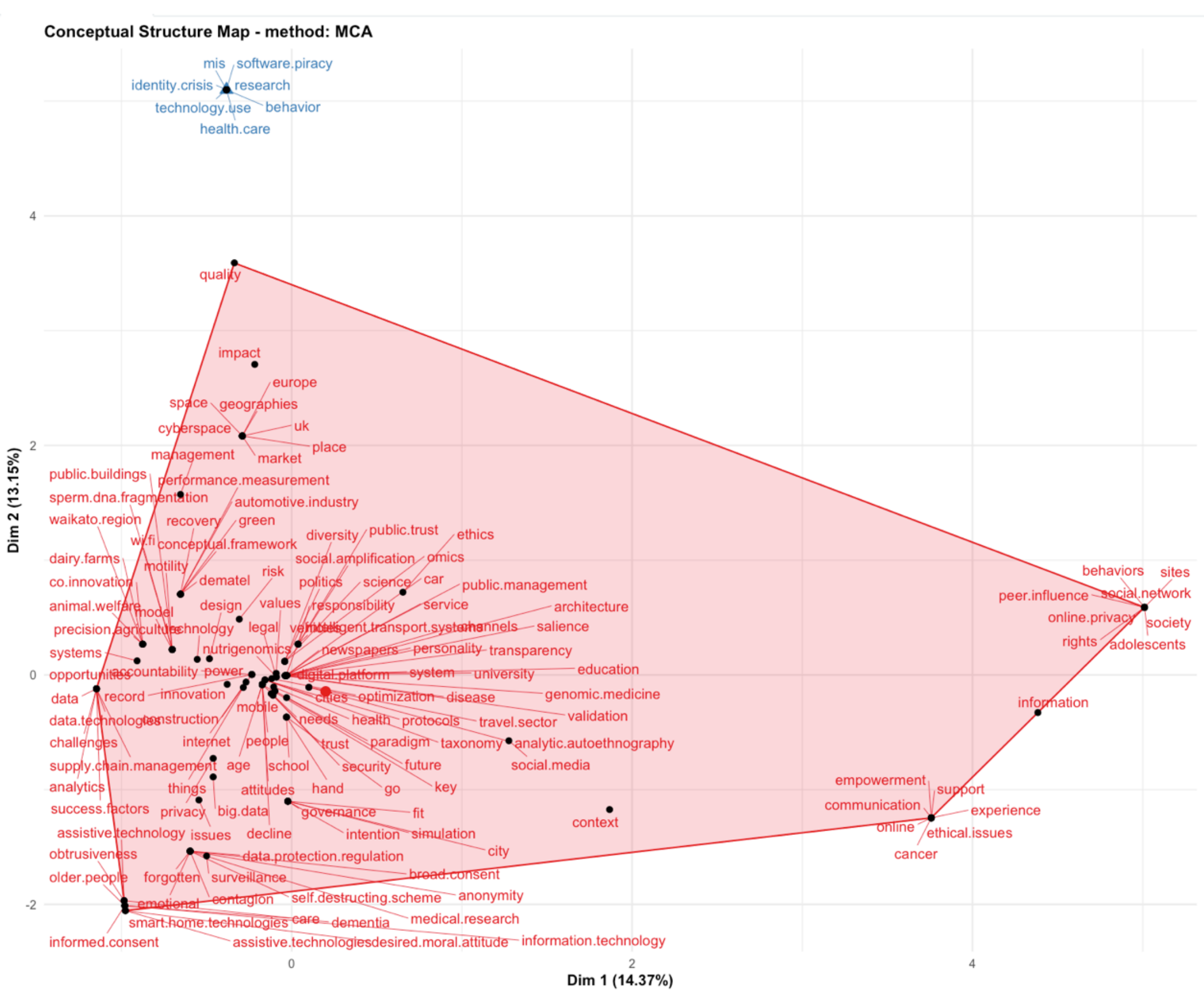

*Figure 2: bibliometric conceptual structure map - factorial analysis*

From the factorial analysis in Figure 2, we can see the topics investigated in different research clusters. Although there is no clear connection between the topics researched in this study, we can see the keywords: medicine, medical research, and care. Which signifies that although there are no related research studies at present, the research community is actively looking at these topics in combination with IoT, new technologies, cyber-risk and ethics.

A separate search on 'internet of things, AND cyber risk', produced 192 records. We designed a three-fields plot (in Figure 3), and we can see a very strong interconnection in the research on IoT and cyber risk, but the keywords related to shared responsibility in health policy are still missing.

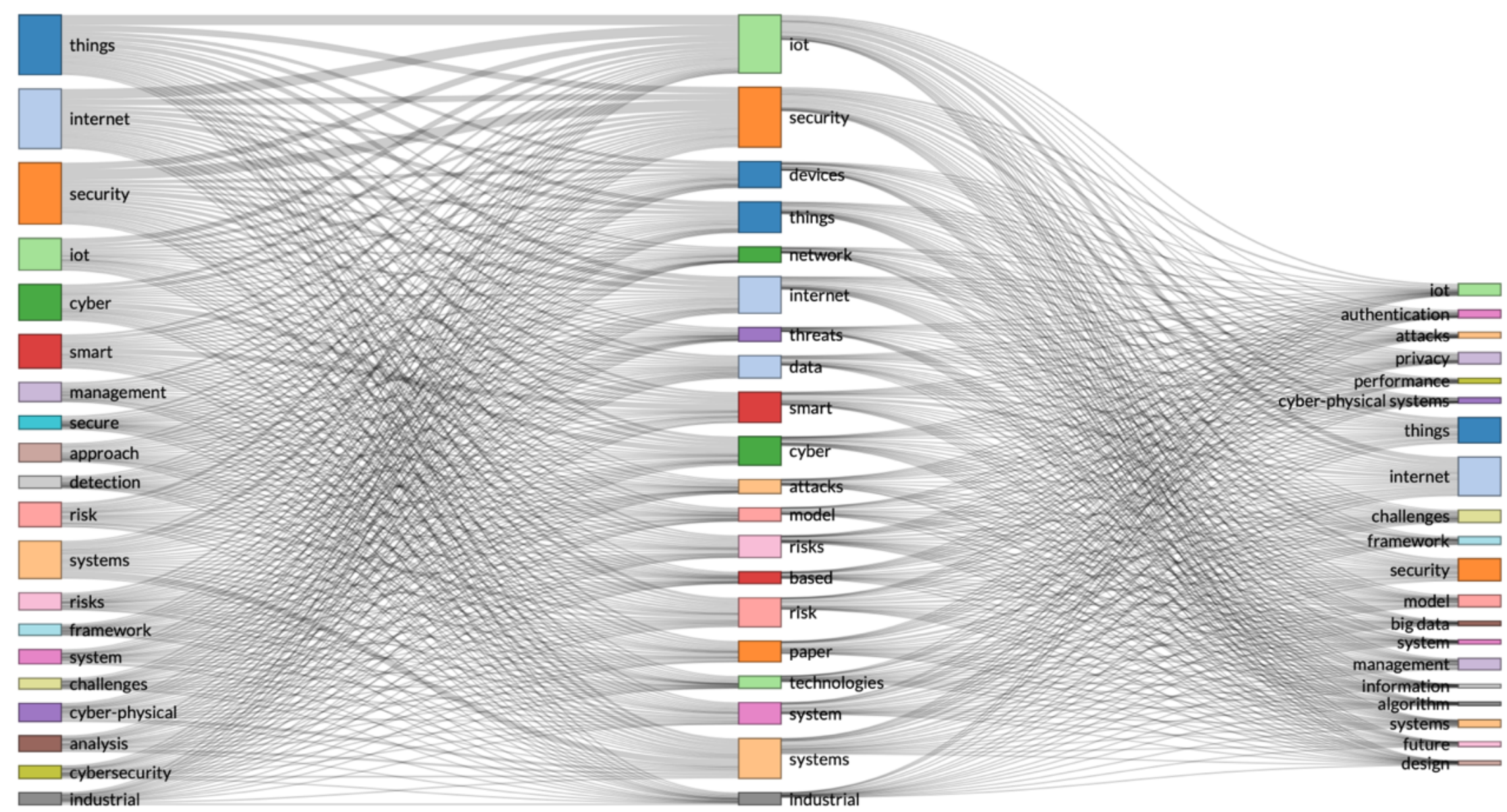

*Figure 3: bibliometric analysis - three-fields plot: keywords from abstract in the middle, keywords from the titles on the right, keywords plus - from all article on the left.*

We designed an alternative three-fields plot (in Figure 4), and separated the field by journal sources, keywords from articles, and most relevant publications (chosen by the R 'bibliometrix' package). Since this data record was on IoT and cyber risk, we were trying to find reference and some connection to health policy. The closes we found was 'healthcare' (see Figure 4).

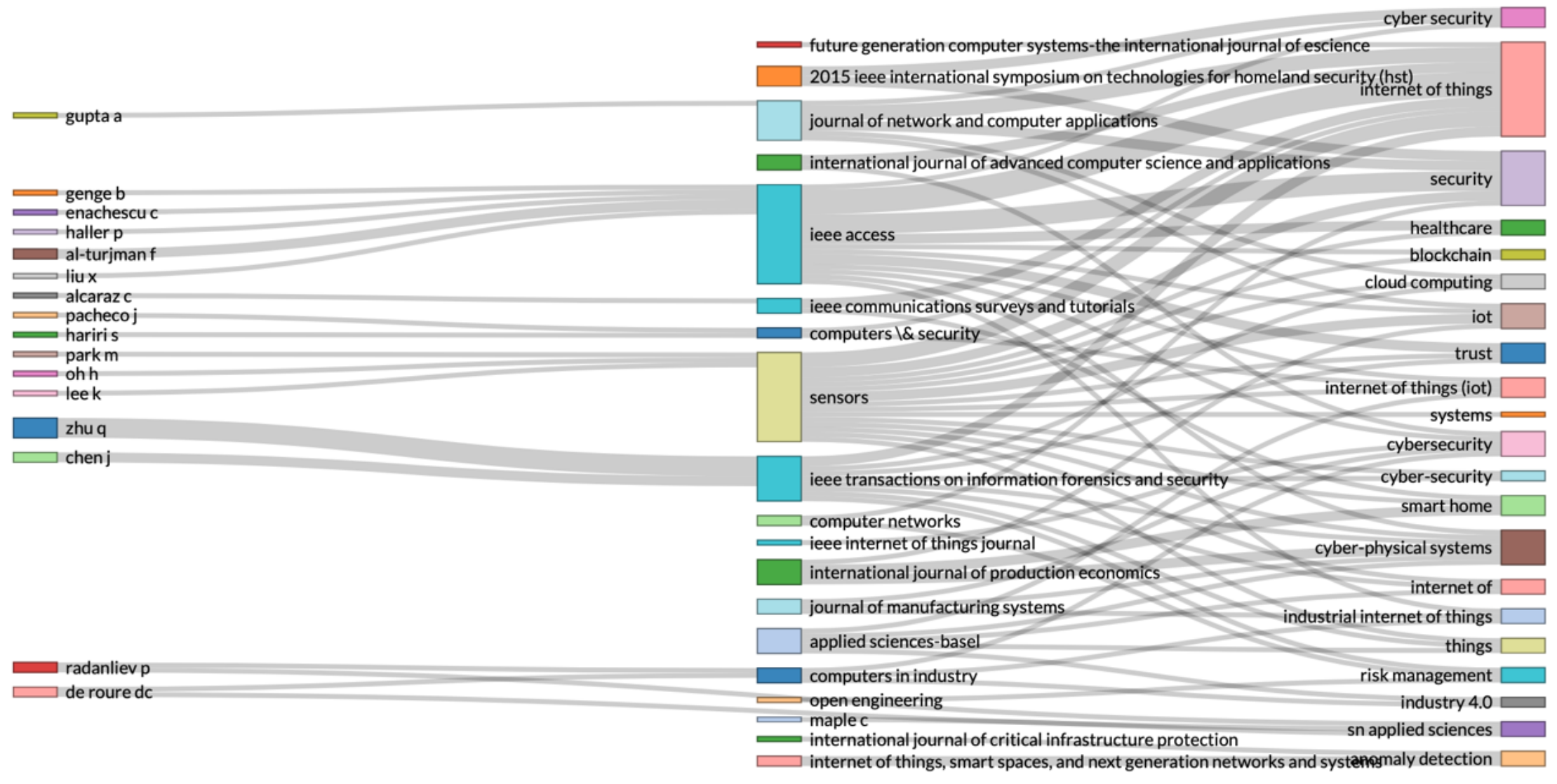

*Figure 4: bibliometric analysis - three-fields plot: keywords from journal sources in the middle, keywords from the articles on the right, most relevant authors on the left (as determined by the software).*

From the bibliometric analysis in Figure 3 and Figure 4, we can see that the topics of IoT, risk, ethics and healthcare are present in research studies, but connection between these topics is still in its infancy. To analyse this further, we designed a co-occurrence network (in Figure 5). In the co-occurrence network, we deleted isolated nodes, and we tried to visualise how different research topics are related.

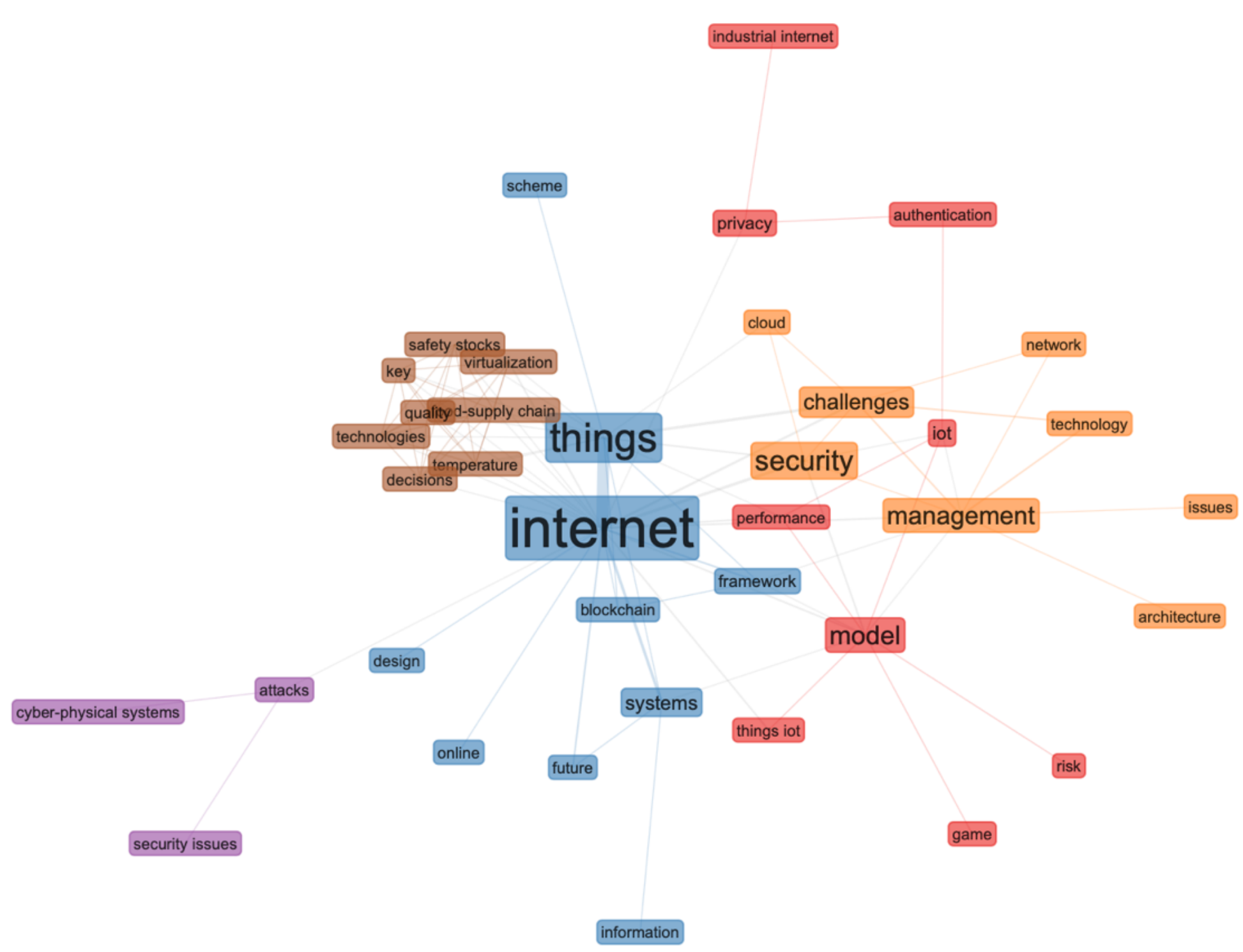

*Figure 5: Bibliometric analysis - co-occurrence network*

From the co-occurrence network (in Figure 5), we can see the research topics in colour coded networks. The healthcare supply chain is in the brown network, security in orange, risk in red, attacks in purple and trust is missing. This analysis confirms that although trust was present in the Figure 4, the topic of trust in this data record was not corelated to the research topics we investigate.

Our final data file to search correlations between ethics, shared responsibility, health policy and IoT, contained 20 records from a separate search on 'ethics, AND cyber risk'. Form this limited data record, when analysed with the R 'bibliometrix' package (Figure 6), we extracted some interesting findings. What this data record shows, is that research on these topics, is strongly represented by the USA, UK and Australia, and predominantly related to military in the USA, and to war in the UK (see Figure 6). There is almost no records of shared responsibility and health policy. We cannot argue that this findings from a limited data record, represent all research that exist. But we need to mention that the search for data records we conducted, was from the Web of Science Core Collection, and it included all historical records from 1900 to 2020 on the topics of 'ethics, AND cyber risk'. We detail our search parameters in the spirit of reproducible research, and we include our data records used in plotting the Figure 6 for other researchers to re-analyse.

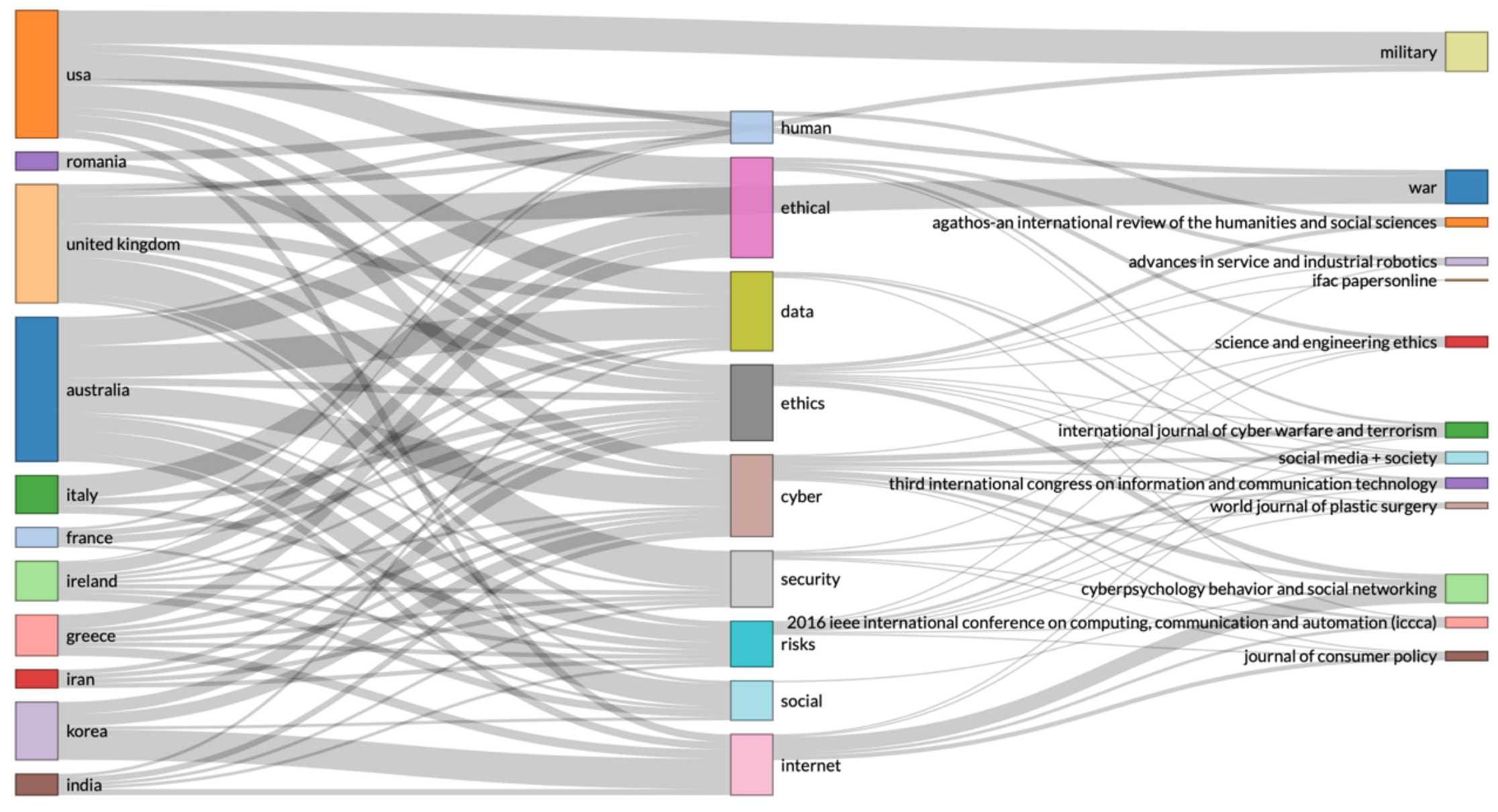

*Figure 6: Bibliometric analysis – three-fields plot from the data record on ethics, AND cyber risk, relating research papers by country on the left, by source on the right, and by common words association from the records abstracts in the middle field.*

If the research parameters are changed to 'ethics, OR cyber risk', this figure will change significantly, but that would represent a bibliographic analysis of records that include either of the two topics and would not represent scientific records that are related to both topics. To describe the difference between this simple different in search between 'AND'/'OR', we conducted a search on the Web of Science Core Collection, and changed parameters to ‘ethics, OR cyber risk’. The search resulted with 155,931 records, which is significantly more, compared to the 20 records on ‘ethics, AND cyber risk’ used to plot the Figure 6. To produce the same three-fields plot with this data record, we could use the R studio. R is designed for analysing big data and can handle such large data record. The Web of Science Core Collection, however, only allows extracting 500 records at a time. To create a data file of these records, we would need to extract 311 separate files, and then merge these files before we can upload the data record in the R ‘bibliometrix’ package. If this data record was considered relevant to ethics, shared responsibility, health policy and IoT, we could have engaged in such labour intense data collection. But such data record would have simply presented a graph of research on wither of the two topics, not research relating the two topics. To analyse the 155,931 records, we used the Web of Science record analysing tool.

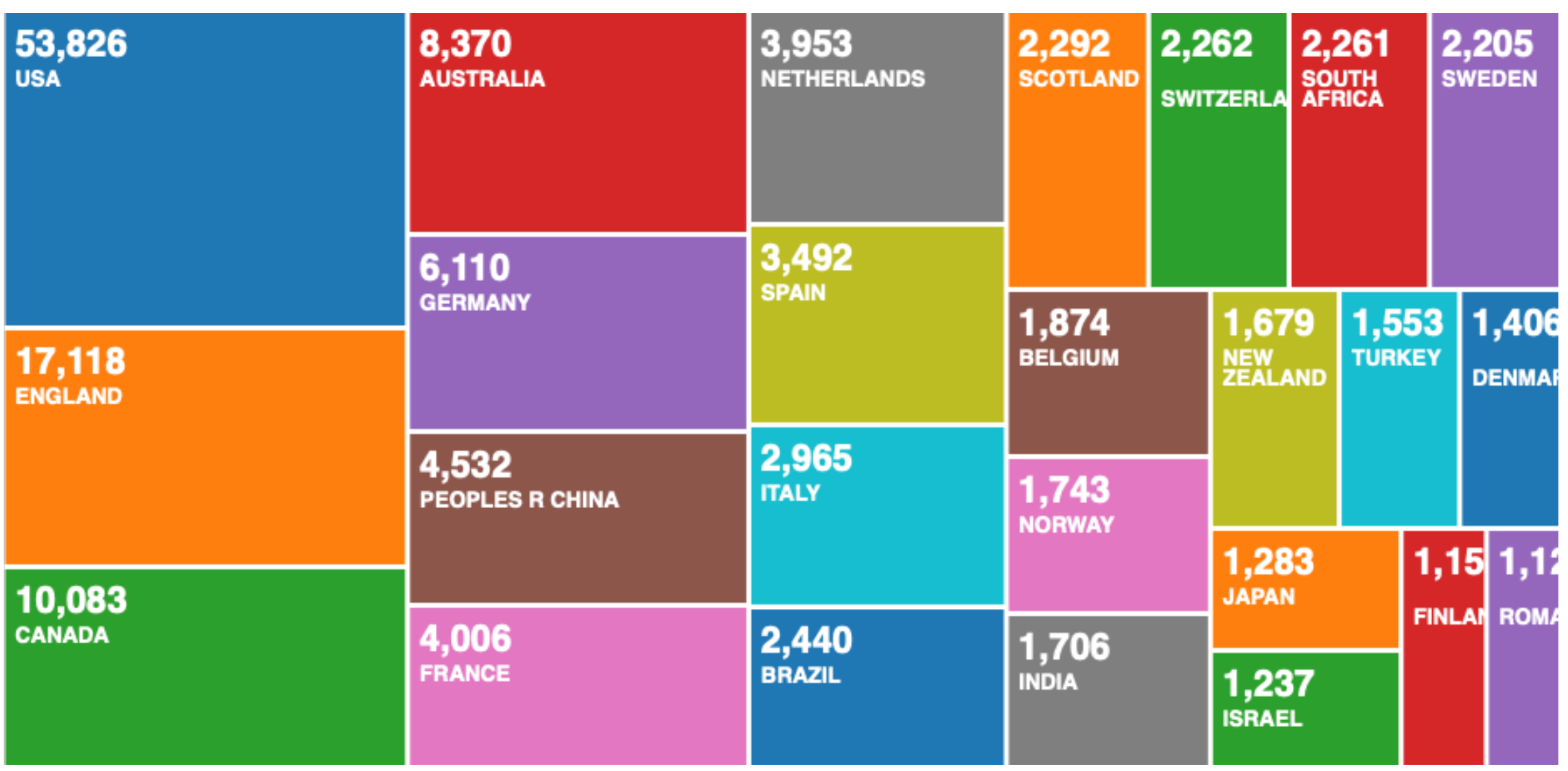

*Figure 7: Country categorisation of 155,931 records on ‘ethics, OR cyber risk’, analysed with the Web of Science record analysing tool*

To identify if the research on ‘ethics, OR cyber risk’ is related to shared responsibility and health policy of IoT systems, or also predominated in the research sources of ‘military’ and ‘war’ as identified in the Figure 6, we also categorised the 155,931 records in research areas (in Figure 8) and institutions (in Figure 9).

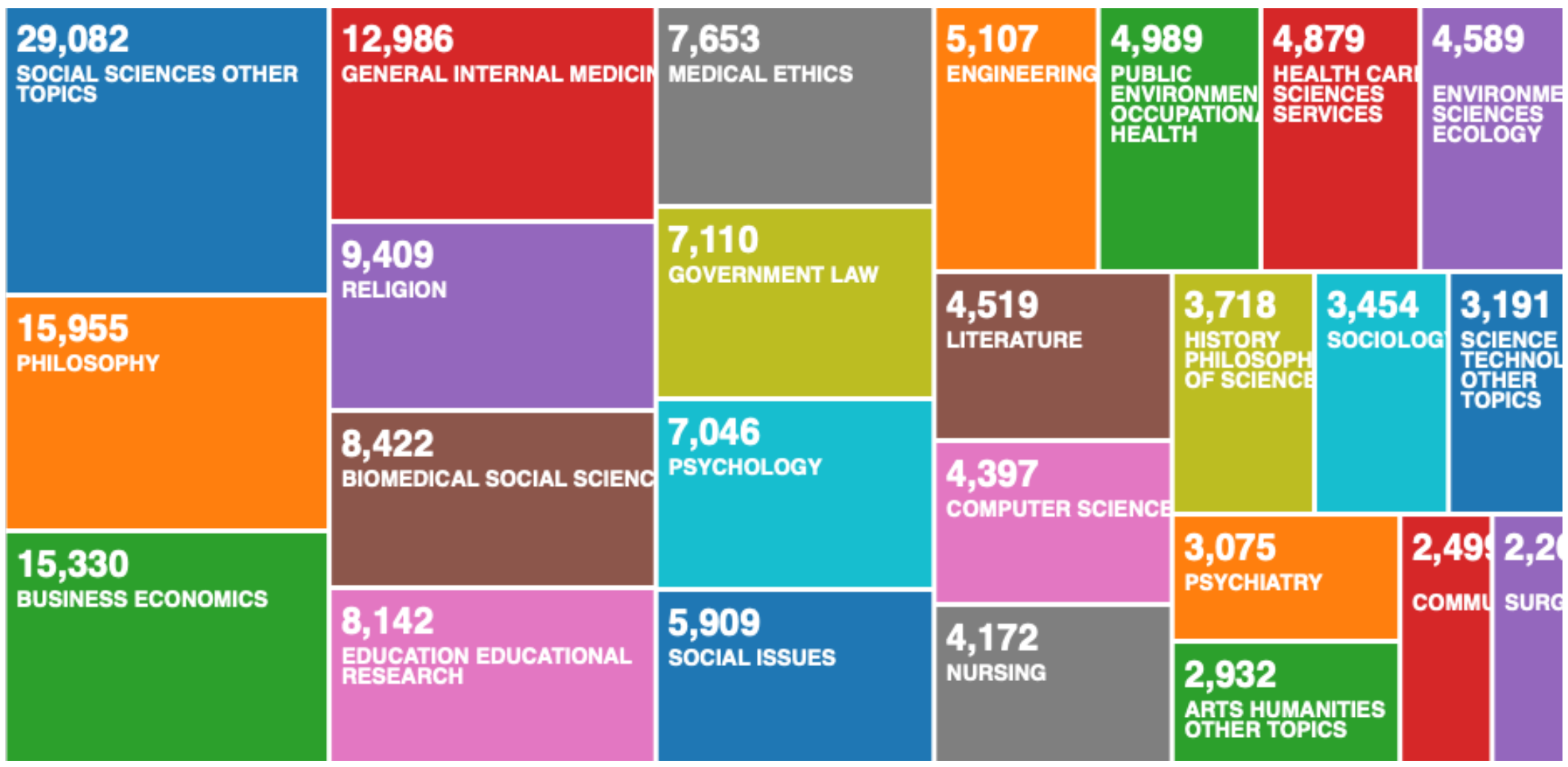

*Figure 8: Research areas categorisation of 155,931 records on 'ethics, OR cyber risk', analysed with the Web of Science record analysing tool*

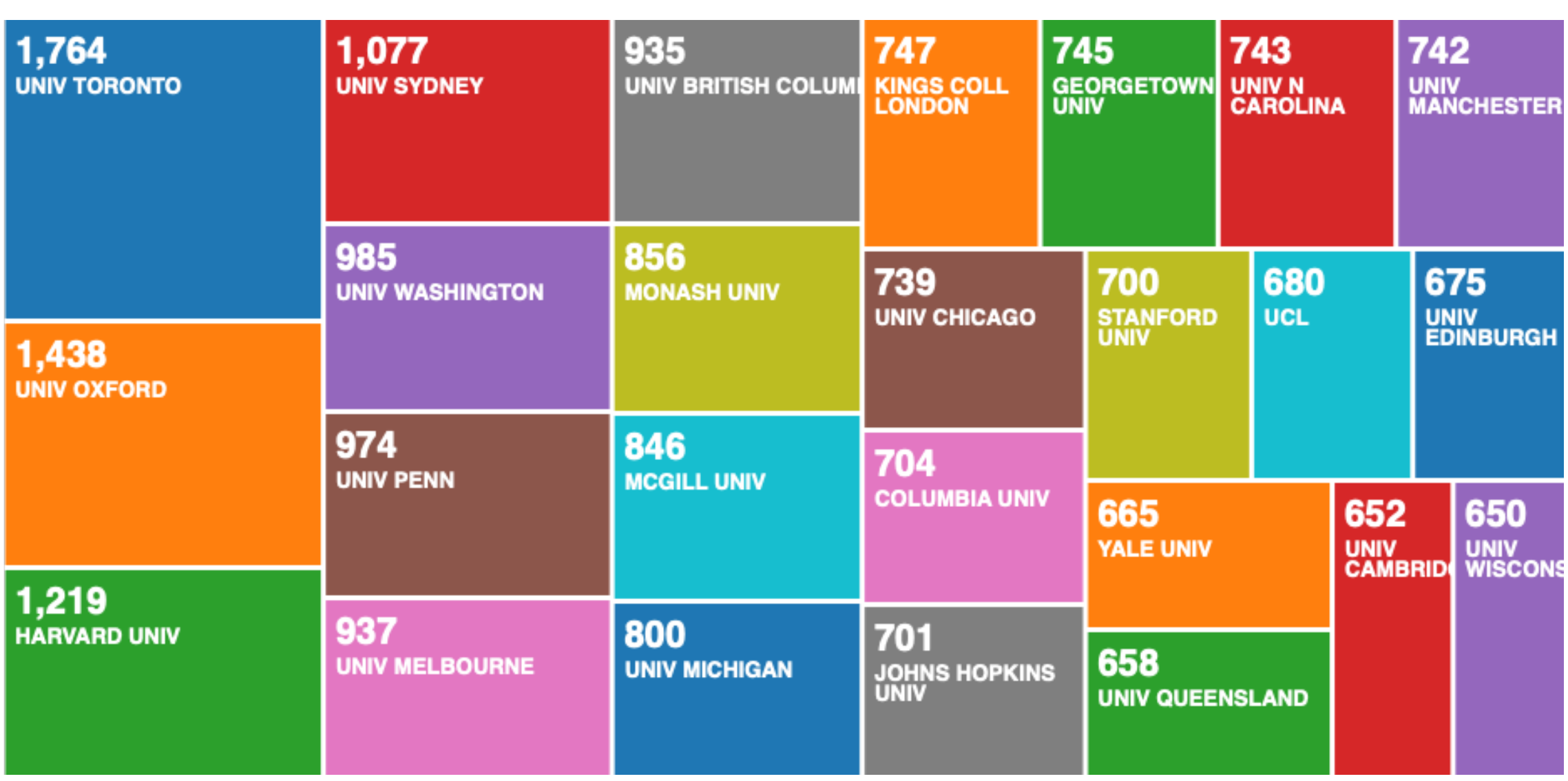

*Figure 9: Research institutions categorisation of 155,931 records on 'ethics, OR cyber risk', analysed with the Web of Science record analysing tool*

What the categorisations in Figure 8 and Figure 9 show, is that scientific research on 'ethics, OR cyber risk', is strongly present in the research areas on social sciences, but there are also a significant number of records on medical ethics and healthcare sciences. The categorisations in Figure 8 and Figure 9 also shows that scientific research on 'ethics, OR cyber risk' is predominately undertaken by academic and not military institutions. Some of these academic institutions are also leading the research in ethics, shared responsibility, health policy and IoT.

The reasons why US and UK are strongly represented in Figure 6, can be explained in the categorisation in Figure 7, which confirms the US and UK (along with Australia) as the leading countries for research on 'ethics, OR cyber risk'. This however, does not explain why research on 'ethics, AND cyber risk' is predominated in the research sources of 'military' and 'war' (see Figure 6). It does however justify our argument that such research is needed in other research areas, which are not related to 'military' and 'war', for example: ethics and cyber risk from the IoT connected devices in health policy.

### *2.1 Explanation on the limitations of our data records*

Since we couldn't find records that included all search keywords on ethics, shared responsibility, health policy and IoT, we had to separate the search keywords. Even before we applied statistical software to these results, we knew that these research records, would not be directly related to shared responsibility and health policy. Instead, we used the data records to visualise the connection between different keywords.

A particularly interesting result emerged when we searched for: 'internet of things, OR ethics, OR cyber risk' (analysed in Figure 10 and Figure 11). We found 199,902 records, which confirmed that these subjects are studied excessively in isolation, or with other research topics. But there is very little research on how IoT technology created new cyber risks, and what are the ethical concerns of this new risk in shared responsibility and health policy for IoT systems.

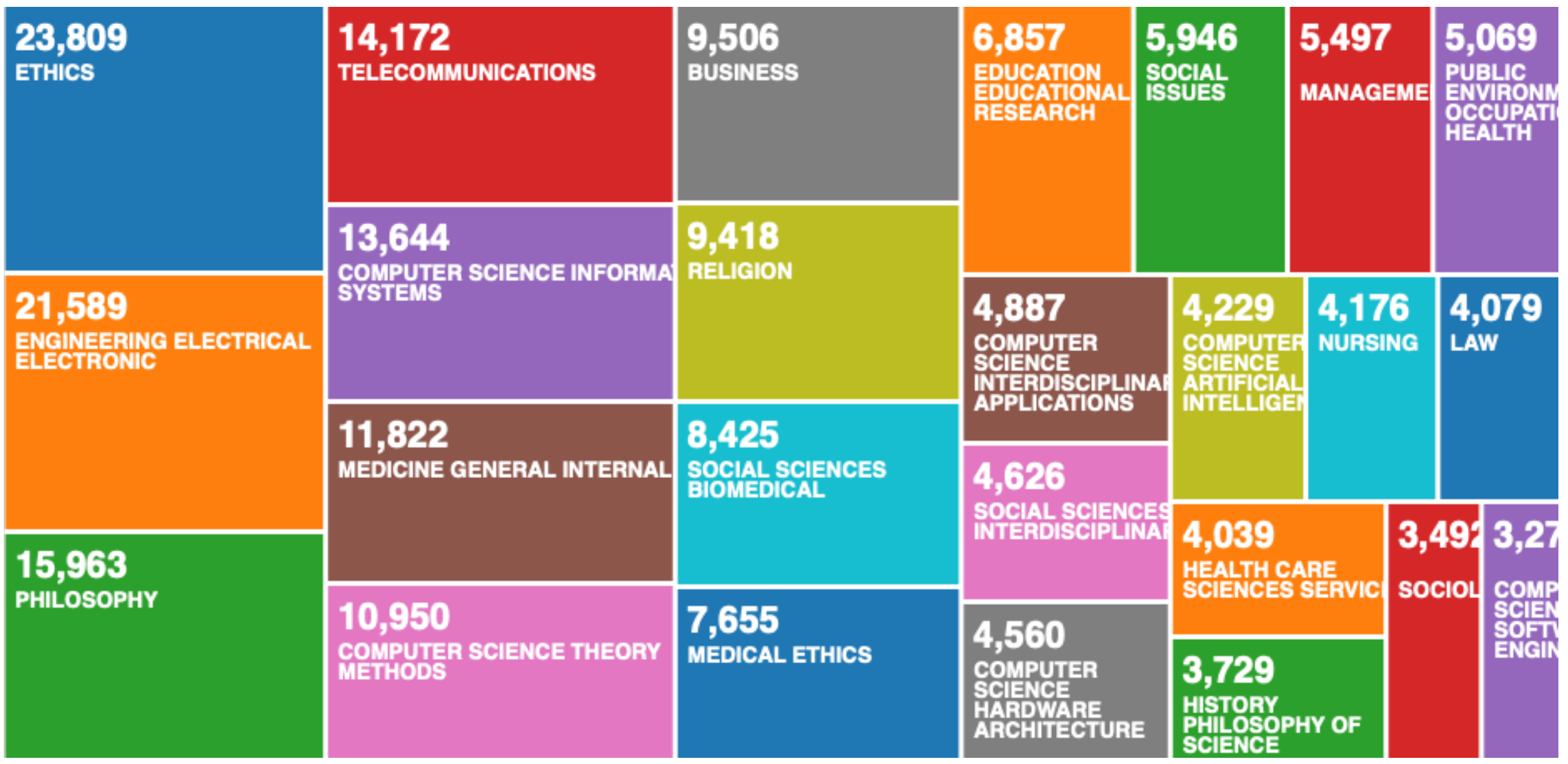

*Figure 10: Scientific research from 199,902 records by field - TOPIC: (ethics) OR TOPIC: (cyber risk) OR TOPIC: (internet of things).*

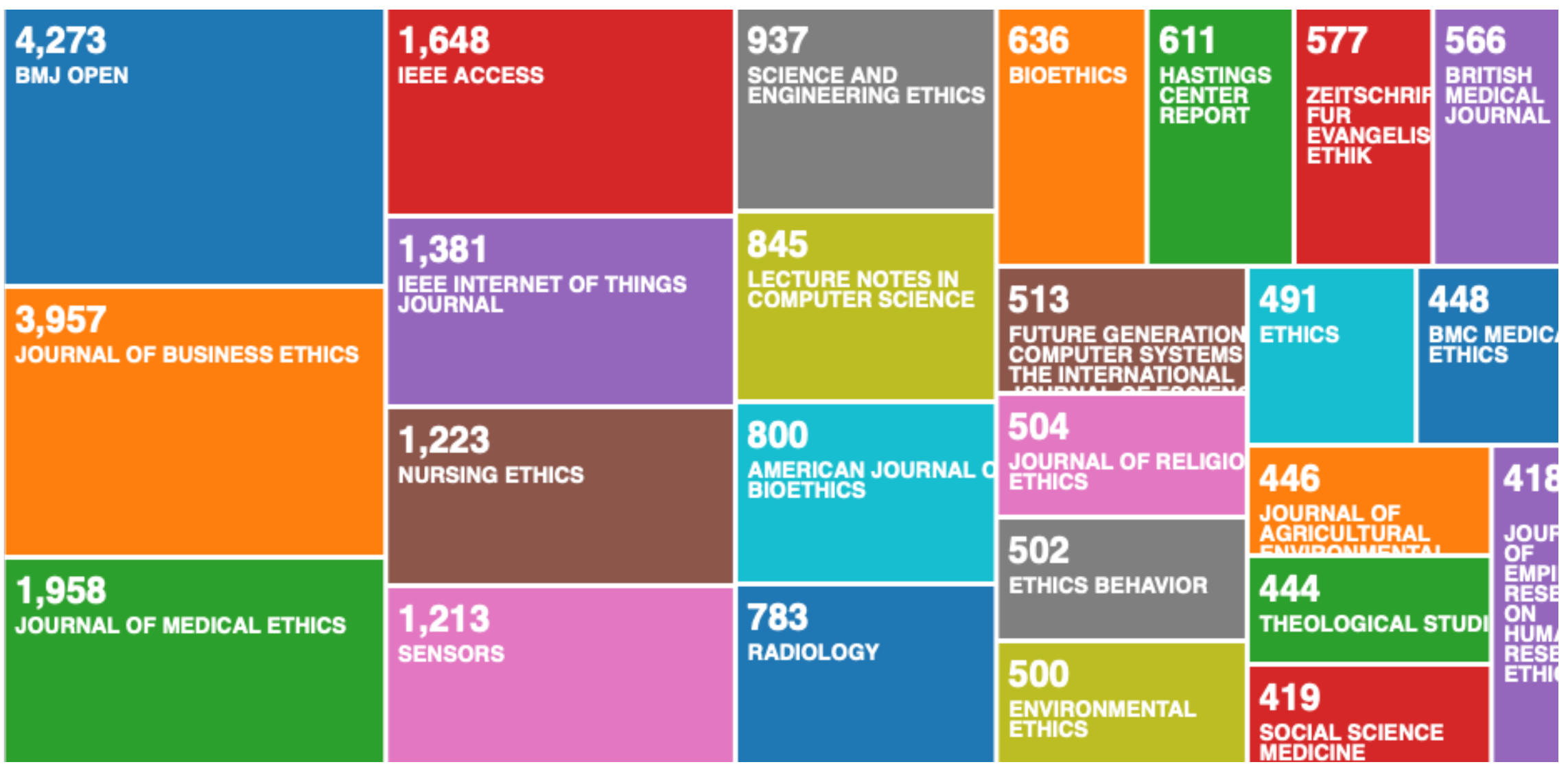

*Figure 11: Scientific research from 199,902 records by journal - TOPIC: (ethics) OR TOPIC: (cyber risk) OR TOPIC: (internet of things).*

## *2.2 Summary of the bibliometric analysis*

In the bibliometric analysis, this paper provides an extensive review of the research on the ethical issues in relation to cyber risk from IoT. Considering that healthcare systems have been studied for much longer than IoT has been in existence, we conducted an extensive review on specific healthcare aspects of IoT in ethics, shared responsibility and health policy. There are fundamental differences between the bibliometric analysis which was conducted with statistical methods, and the case study, which was conducted with epistemological analysis. The objective of the case study was to relate a multiverse of keywords on ethics, shared responsibility, health policy and IoT in an epistemological framework that connects the topics of ethics and shared risk.

The bibliometric analysis showed that, there are 'white spots' in scientific research on the topics of IoT and the ethics of shared cyber risk in health policy. This is made clear in our search for data records and analysis on the topics of 'internet of things, AND ethics, AND cyber risk'.

Nevertheless, this is not an attempt to persuade the reader that there is no scientific research on these individual topics, as seen in our search results (199,902 records in Figure 10 and Figure 11) for data records on 'internet of things, OR ethics, OR cyber risk'.

Our conclusion from the bibliometric analysis is that research on 'internet of things, AND ethics, AND cyber risk' is limited. But to say that little research has been conducted on the topic of 'ethics OR cyber risk' (155,931 records in Figure 9) or on shared risks form IoT system in health policy, could be misleading. such research does exist, as described in (Figure 10 and Figure 11). However, the statistical analysis we conducted shows that such research is not connecting the topics of shared responsibility and health policy with the research on IoT, ethics, and cyber risk.

Finally, in the bibliometric analysis we focused primarily on the topics of ethical research on IoT and cyber risk, and we conducted a separate case study on shared responsibility and health policy when Covid-19 was in its infancy. Hence, we distinguished in our analysis between ethics, shared responsibility, health policy and IoT, prior the emergence of Covid-19. Although the Covid-19 emerged in late 2019, the research on Covid-19 started emerging in 2020 and the adoption of IoT technologies in health policy has taken much longer. We argue that since some of the most prominent literature on ethics, shared responsibility, health policy and IoT are from prior to the Covid-19 emergence, the use of modern technologies is not considered as strongly in the development of those studies. Therefore, the case study is used to extract knowledge from established technological approaches to change, that are designed prior to the emergence of Covid-19 and the integration of IoT in healthcare, and its associated cyber risk. We then use epistemological approach to use that knowledge in our study.

## 3 Case study as a research methodology in combination with epistemology

An epistemological framework in this paper is defined as a process that derives with insights from existing knowledge and represents a method for providing guidance for applying knowledge in practice.

The epistemological approach is used here as a method that provides clarity in blurred subjects. The reason ethical risks in IoT-enabled health policy are considered a 'blurred subject' in this study, is the bibliographical analysis in the previous section. Shared responsibility represents a healthcare system where all participants are working towards a common goal, but in Covid-19 management systems, individual healthcare participants are interested in managing Covid-19 and no other illnesses. Healthcare system do not share profits, but are designed to lower cost and deliver service, among many other functions; each participant in the healthcare system is responsible for managing their own department, hence, the primary concern is with eliminating that specific medical problem. However, for the system to operate, participants should be concerned about the shared responsibilities. This raises questions on who is responsible for investing in cybersecurity of medical devices. Is cybersecurity on shared IoT technology an investment in healthcare, or an expenditure? If intelligence data is collected from patients, who owns that data? Who is responsible for compliance with regulations? Such blurred subjects are frequently researched with more flexible qualitative methods [4], [5], such as case study research.

The research methodology adopted for our article's research is qualitative and explorative in nature using primary and secondary data resources. The data is synthesised using the case study research methodology, using qualitative primary and secondary resources and categorising emergent ethical concepts into themes.

Academic literature is consulted extensively, and then a case study research methodology is applied to build an epistemological framework from the literature review [6]. The case study method is applied with the ethnographic and discourse approaches [7], for the construction of scientific theory [8] on the ethics of shared cyber risks.

The methodology we selected for ensuring validity of the findings is focused on applying qualitative research techniques that included open and categorical coding to analyse and categorise the qualitative data [9]–[11]. The qualitative data is collected from journals on the topics of shared risk from IoT technologies, and cyber risk in the IoT. Open coding was used to validate that there is a reliable representation of the data collected, while categorical coding subsequently was applied to recognise the profounder concepts in the data [12]. Discourse analysis is applied to evaluate and interpret the connotation behind the explicitly stated approaches [11], along with tables of evidence [13] and conceptual diagrams [14] to present graphical analysis.

### *3.1 Case study*

The case study begins by requesting the participants to define an overall health, care, and health technology objective as an ethical approach for applying IoT technology for managing Covid-19. Given the difficulties for conducting in person case study interviews and workshops during Covid-19 lockdowns, the case study is performed with a limited number of most crucial participants, and these limitations with participants size are enhanced with bibliometric and statistical analysis of data records from the Web of Science databases. To overcome the limitations, a literature review is also performed on this topic and the results are analysed with the case study method (see summary map in Table 1). The pool of 20 participants interviewed were proportionally representative of different levels of seniority and included Senior Directors, Principal Engineers and Risk Managers. The statistical data of these interviews in analysed, condensed, anonymised and presented (see Figure 12-16) with the use of the Pugh controlled convergence method [15]. The initial participants were selected through convenience sampling. We conducted semi-structured interviews, and we gained informed consent for the study. Only part of the interviews was predetermined in the initial selection and the remainder were chosen based on the development of the case study research. This process corresponds with existing literature [16]. The length of the interviews was predetermined as one hour, and the data collected was transcribed and categorised with aims to investigate the ethical relationship between the notion of IoT and the associated shared responsibility. Following the advice from recent literature on ethical assessments of Covid-19 healthcare strategies [17], the aim of the analysis was to identify the ethical ideas behind the statements and to relate in Fig. 12, with the categories of 'nominal' or 'desired' ethical strategy [17].

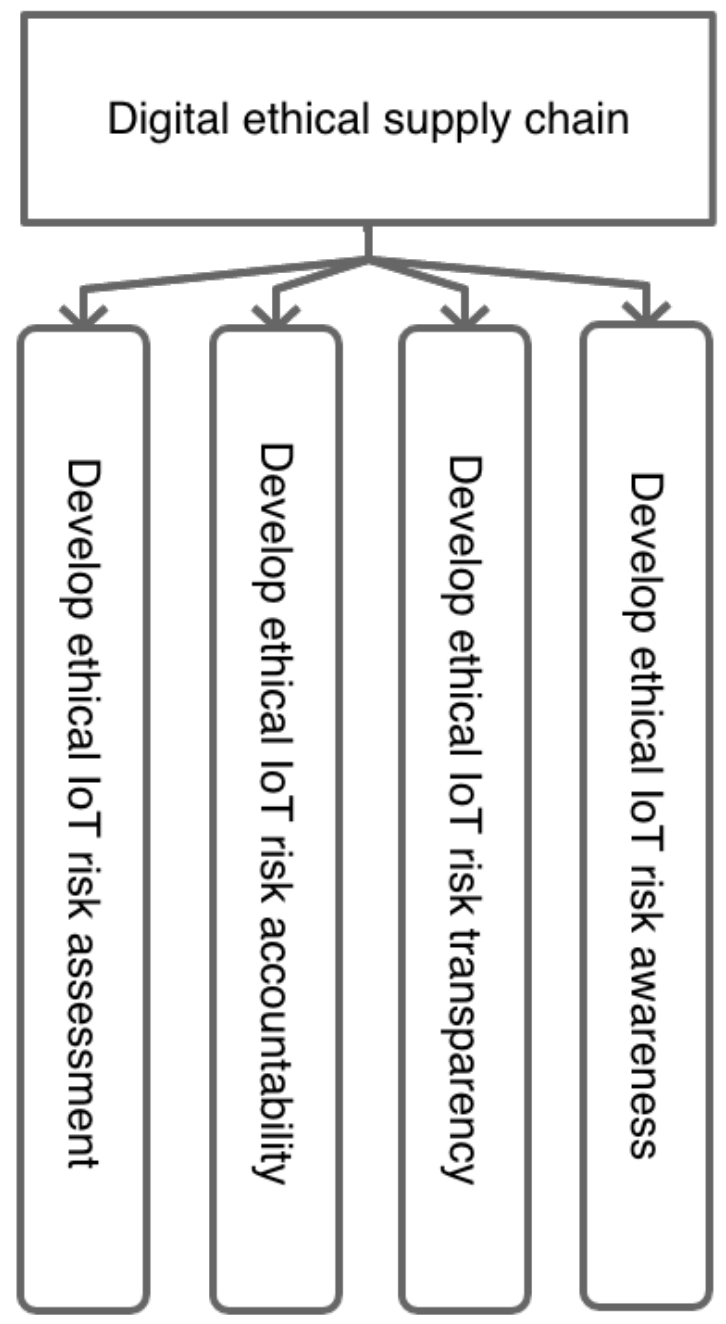

Fig. 12: Categorising an ethical 'nominal' strategy for using IoT technology in health policy for shared responsibility

This process of categorising concepts clarified individual levels of the ethical assessment of cyber risk in shared healthcare systems for health policy for shared responsibility. The categorising of the ethical 'nominal' strategy enables clarity in the individual levels of IoT technological integration. To categorise the second level, the case study embodied a process of ideas and concepts conceived as an interrelated, interworking set of objectives that enable the development of systematic understanding of ethics and the cyber risks in health policy for shared responsibility.

Directive, conventional, and summative analysis were applied to analyse and categorise the ethical concepts emerging from the interviews. The Fig. 13 outlines sample themes illustrating the stated categories and subcategories relating the concepts with the epistemological framework. The process was built upon the open and categorical coding methodology (Fig. 13) to evaluate the epistemological framework.

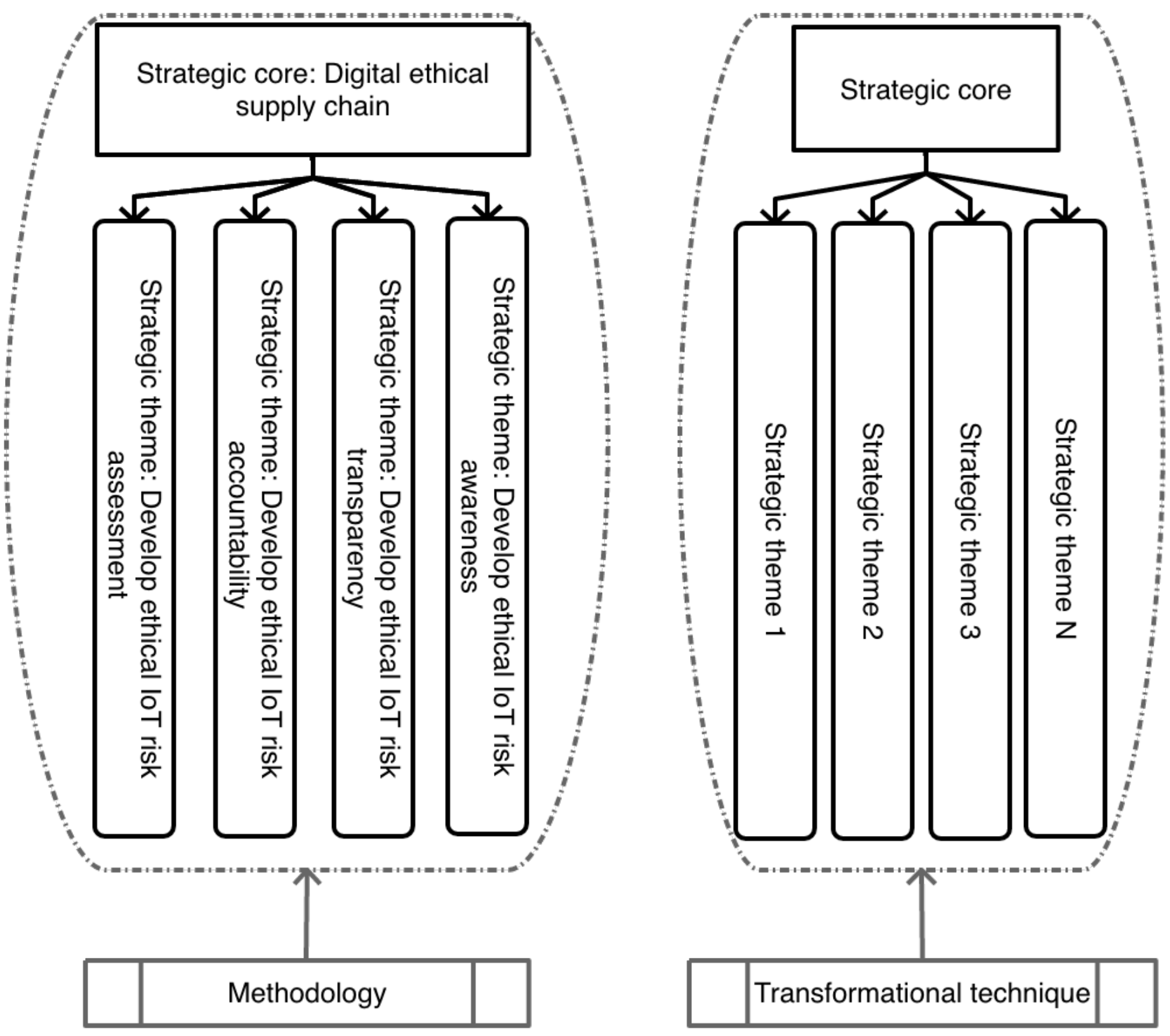

Fig. 13: Open and categorical coding methodology for applying the epistemological framework in health policy for shared responsibility

The diagram in Fig. 15 presents graphical analysis of the framework as a transformational technique that can be applied as a methodology in practice. The graphical analysis represents the step-by-step description of the ethical 'nominal' strategy, emerging from the findings from decades of research on ethics, shared responsibility, health policy and IoT. The process in Fig. 12 and Fig. 13 enables identifying and relating the functional themes behind individual strategic themes (as described in Fig. 12). To verify the design, open coding was applied to provide a 'nominal' representation [18], while categorical coding was applied to identify the profounder concepts [12].

The process describes how open and categorical coding methodology was applied to defining a strategic core, from where the strategic themes emerge as guiding ethical themes. The themes are supported by functional themes, representing ethical action points in the ethical roadmap.

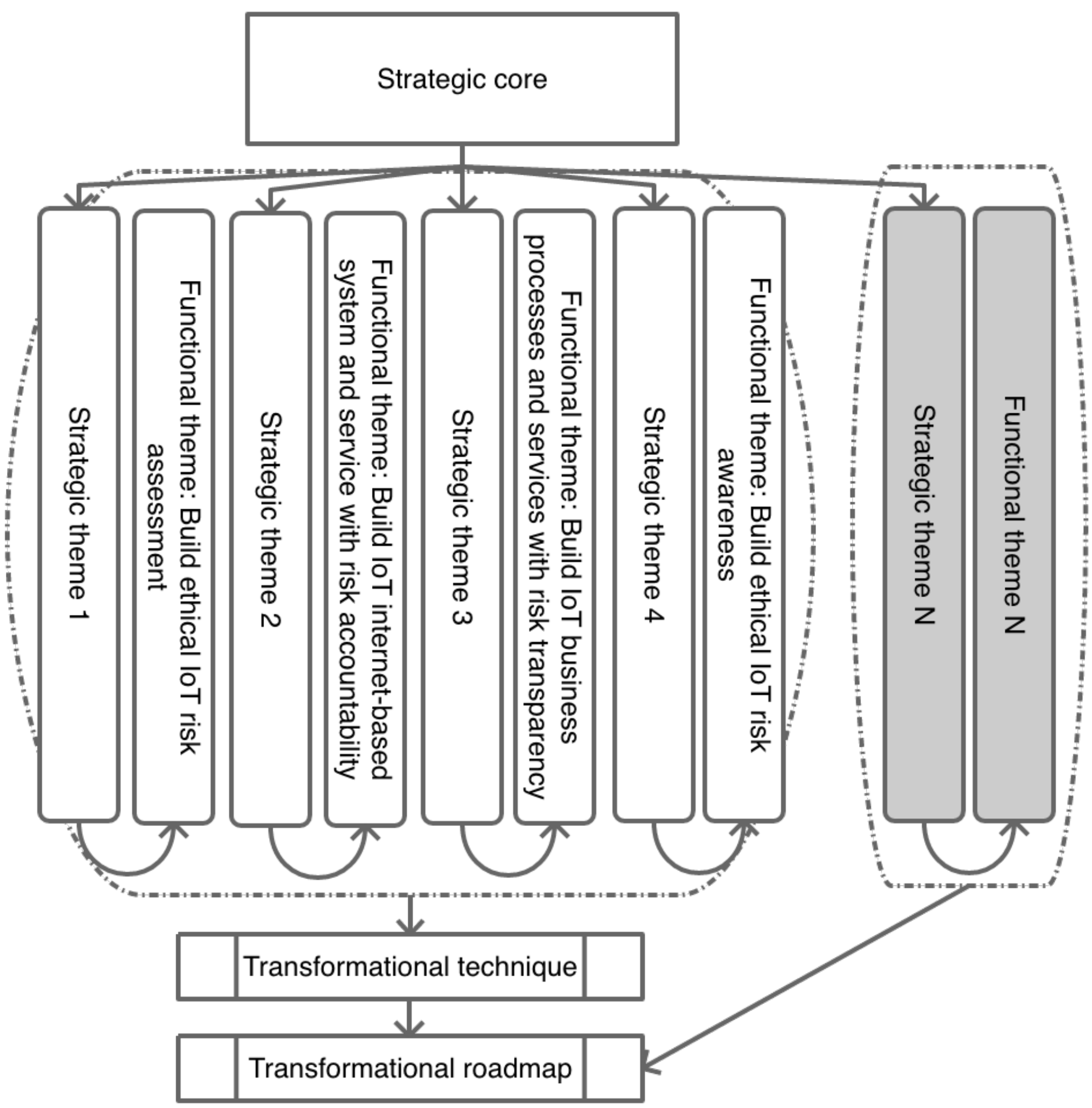

Fig. 14: Populating the design process for the 'nominal' part of the epistemological framework on awareness, transparency, accountability, and ethical assessment of shared risks in health policy for shared responsibility in IoT systems.

The roadmap as expressed (in Fig. 14) represents categories of statement. These statements are related to populate the categories of the ethical assessment and continue populating the roadmap (Fig. 14) with further ethical themes. The roadmap is designed from the scaffolding in the epistemological framework. The scaffolding utilises the knowledge from the epistemological framework and applies the knowledge for ethical assessment of cyber risk in individual operations from shared healthcare infrastructures. The graphical analysis in Fig. 14 and Fig. 15 represent the design process for building the complete epistemological framework, required for ethical assessment of the shared cyber risks emerging from IoT technology. To build the complete epistemological framework, we followed the above-described process, and the case study research collected the emerging concepts to categorise and populate the remaining themes.

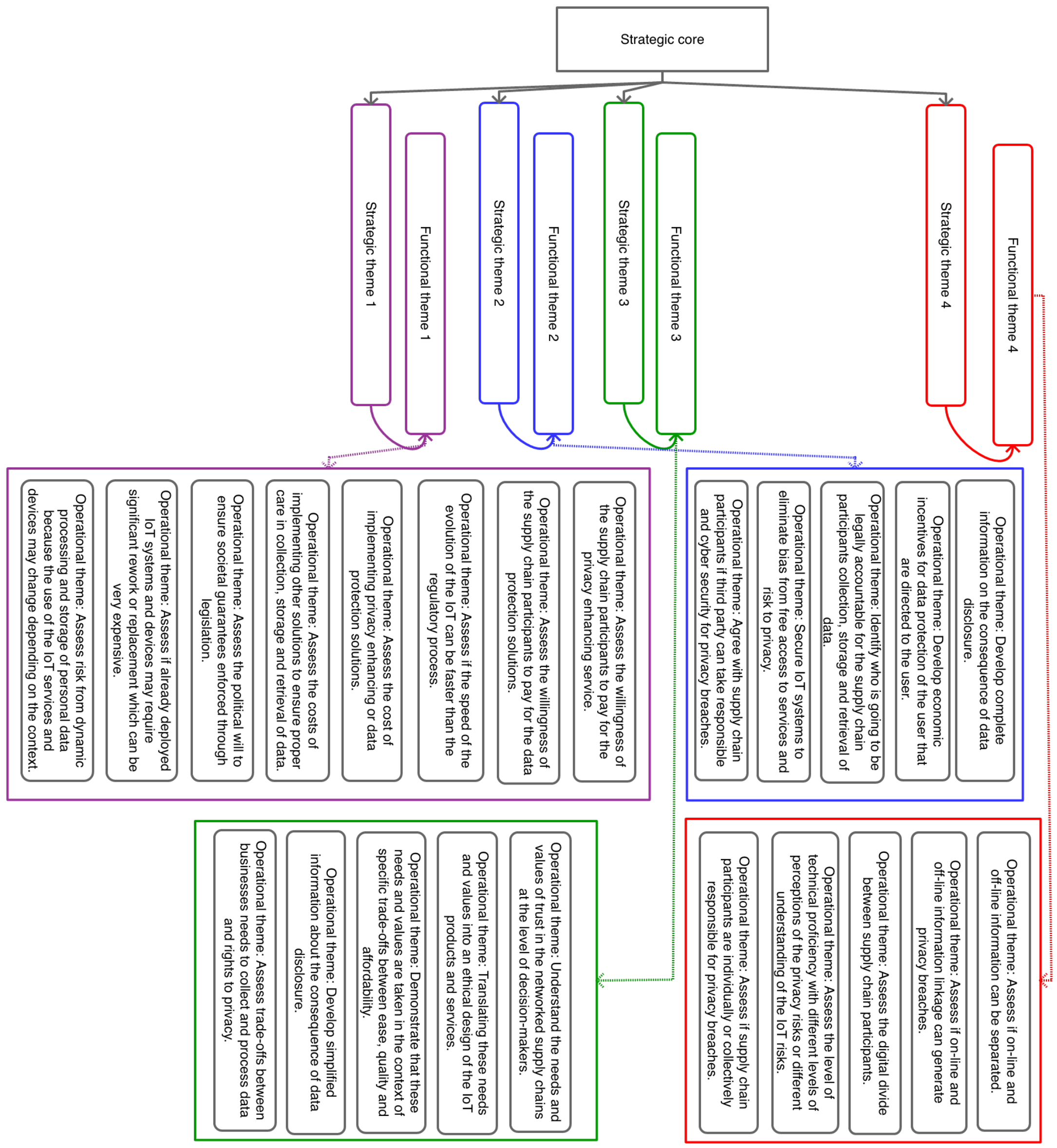

Fig. 15: Populating the design process for the 'executed' part of the epistemological framework on awareness, transparency, accountability, and assessment of shared risks in health policy of IoT systems.

The graphical analysis in Fig. 15 present the case study design process for building the 'executed' part of the epistemological framework. Detailed summary maps and tables with descriptions of the rationale for each task in Fig. 15 and the related references can be found in our project report, which is available as a preprint of earlier work. To avoid duplication of work, we exclude the description tables from this article, but this open-source information would be useful for the readers to understand better the main risks. The design indicates the stages of development of IoT technologies and enables awareness, transparency, accountability, and assessment of the cyber risk emerging from individual activities related to health policy of shared responsibility in IoT systems. The final stage of the epistemological framework is populated in Fig. 15 with the operational themes segregated into subcategories of operational tasks. To populate Fig. 15 we developed a description table that is presented as summary map of literature (Table 1). The summary map in Table 1 presents the most critical areas for discovering cyber risks and taking preventative measures against such risks.

| Operational themes | References: |
| --- | --- |
| Electronic and physical security of real-time data. | [19]–[21] |
| Cybersecurity requires information assurance and data security, protection for data in transit from physical and electronic domains and storage facilities. | [22]–[24] |
| Asset management and access control are required for granting or denying requests to information and processing services. | [25] |
| A process is needed to address novel vulnerabilities caused by life cycle issues, diminishing manufacturing sources, and the update of assets. | [26] |
| IoT cyber risk management requires anti-counterfeit and supply chain risk management to counteract malicious supply-chain components modified from their original design to enable disruption or unauthorised function. | [27], [28] |
| Source code access should be limited to crucial and skilled personnel. This can provide software assurance and application security and may be necessary for eliminating deliberate flaws and vulnerabilities. | [29] |
| Security should be supported with forensics, prognostics, and recovery plans, for the analysis of cyber-attacks and coordination for identifying external cyber-attack vectors. | [26], [30], [31] |
| An internal track and trace network process can assist in detecting or preventing the existence of weaknesses in the logistics security controls. | [32], [33] |
| A process for anti-malicious and anti-tamper system engineering is needed to prevent vulnerabilities identified through reverse engineering attacks. | [34] |

*Table 1: Summary map of operational tasks*

The operational themes in Fig. 15 are linked with emerging operational tasks (in Fig. 16). Identically to the process described previously, the case study is used to collect the qualitative data, and to categorise the propounding concepts in the data.

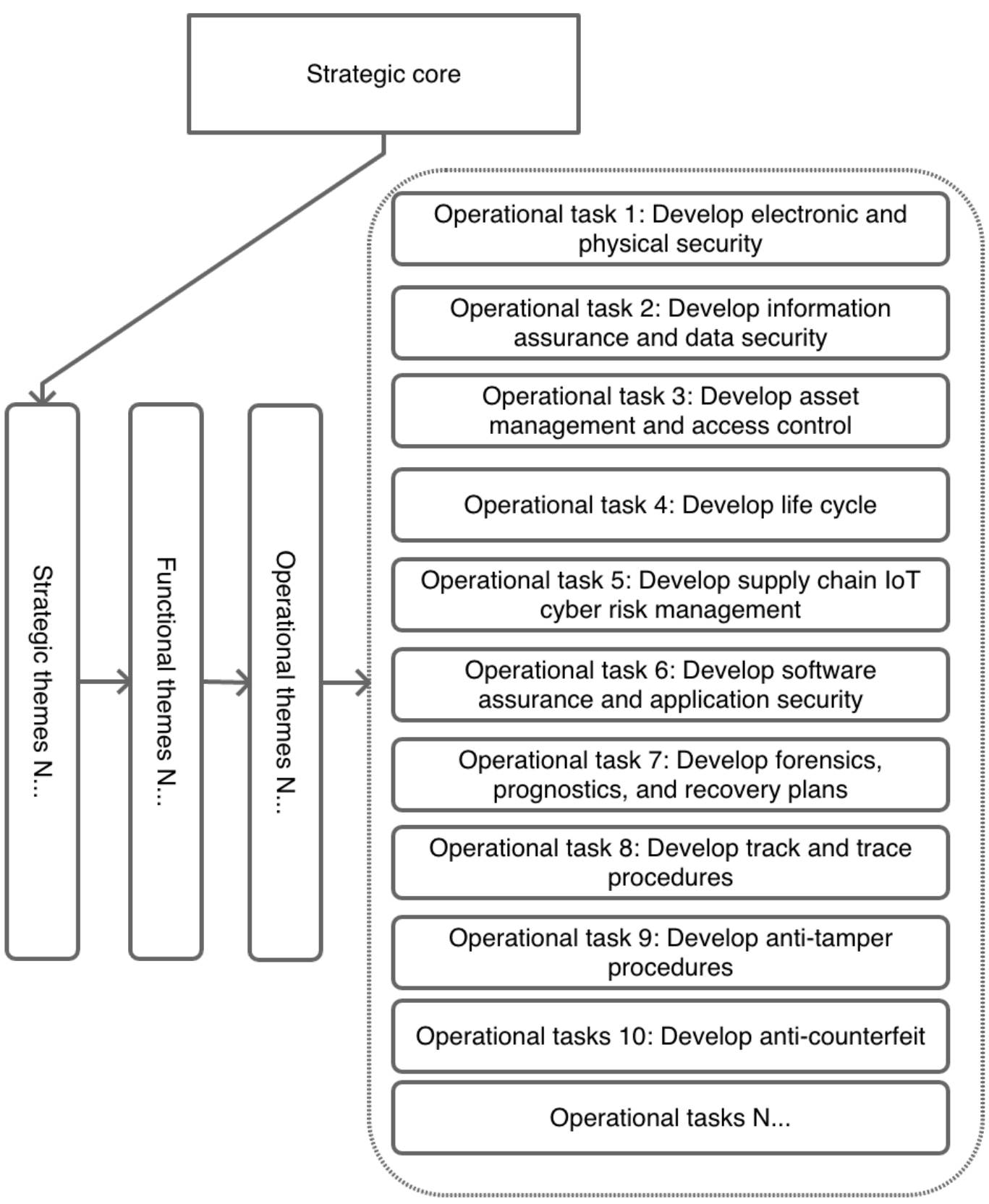

Fig. 16: Populating the final 'operational' step of the epistemological design process for the 'executed' part on awareness, transparency, accountability and ethical assessment of IoT cyber risk in shared infrastructure for health policy of shared responsibility.

This design process (Fig. 16) for cyber risk provides a graphical guidance with themes and subcategorised tasks for ethical assessment of IoT cyber risk in health policy for shared responsibility of IoT systems. The epistemological design enables practitioners to assess the strategic, functional and operational cyber activities and provides a guidance map for ethical cyber assessment.

While the IoT-enabled healthcare adoption requires ethical reference point, the existing healthcare models lacked clarification on shared cyber risk from IoT technologies. By comparing existing models with a case study research with industry participants, the final

'operational' step of the epistemological design in Fig. 16 demystifies this. The ethical design clarifies the required digital capabilities and IoT digital technology in the operational levels of the digital healthcare design. The epistemological design for assessing shared risks, provides practitioners with a step-by-step guidance on how the ethical assessment can be applied to other medical services.

In Fig. 17, the entire epistemological framework for ethical assessment is simplified in one generic diagram. This enables researchers and practitioners to compare the epistemological framework, with their existing process for shared responsibility of IoT systems. The generic design outlines a new approach for ethical assessment of shared cyber risk. The process described in this paper is specific for shared risks from IoT systems in health policy. However, the design process is also generic and could be applied for other types of shared cyber risks.

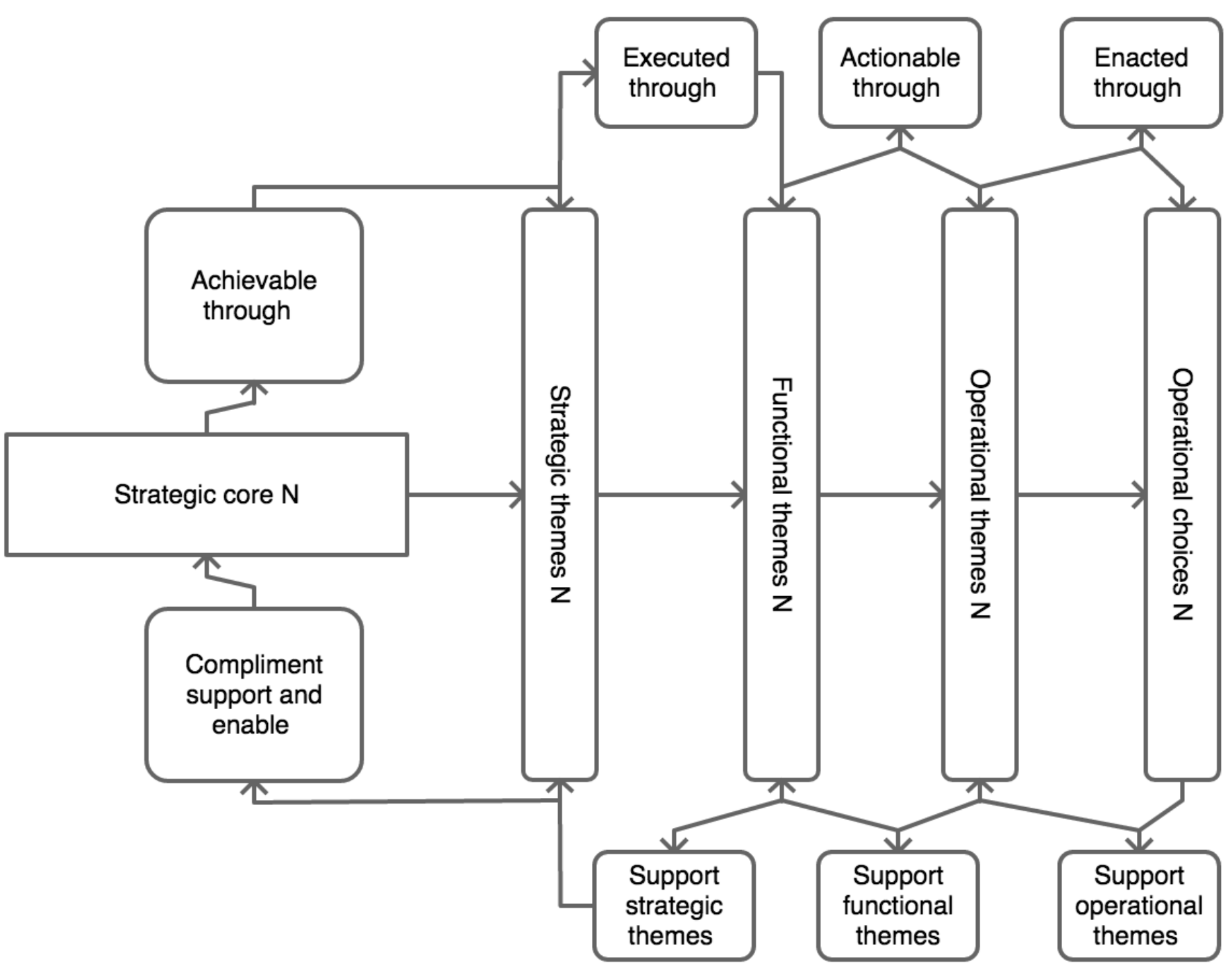

Fig. 17: Generic epistemological framework for ethical assessment of cyber risks from shared infrastructure in IoT systems

The epistemological framework in Fig. 17 pursued common terminology, approaches, and ethical assessment standards, while capturing the best practices for shared healthcare elements in medical IoT systems. The findings can be applied as guidance for academics and practitioners for ethical assessment of shared risks.

### *3.2 Execution of tasks*

The execution of the tasks is one of the crucial points of relevance for future researchers and practitioners applying the epistemological framework. The key to the execution is in the wording of the strategic, functional, and operational themes, and the operational choices. The framework presents a hierarchical structure in the wording, where higher ranked themes present more nominal (descriptive) themes. While the lower ranked themes and choices present imperative statements in the wording. For example, in the Fig. 12 the example of a 'strategic core' is a statement 'Digital ethical'. The statement represents an idea as a vision of something to achieve, but it does not detail the process. An example of a 'strategic theme' from Fig. 13 would be 'Develop ethical IoT risk awareness'.  The wording tells you what to do, but it does not explain how this would be achieved.

An example of a 'functional theme' in Fig. 14 is 'Build ethical IoT risk awareness'. The wording tells you how to do the task, but it does not detail the steps. A good example of a 'operational theme' from Fig. 15 is the theme 'Develop complete information on the consequences of data disclosure'. The wording explains the steps that need to be completed for achieving the goal of the functional theme, which is in a higher ranking, but fails short of actionable task that could enable the theme. The final level of the hierarchy is the 'operational choice' or in the second wording 'operational task', where activities are stated as imperative statements. This level represents the operational activity, which is a choice that each company tasks for execution. One example from Fig. 16 is the task 'Develop forensic, prognostic and recovery plans. The wording of this level is in an imperative form, namely, it tells the organisation what to do in a form of an order. The activity must be clear and executable, otherwise, it would belong to the higher level where themes are more abstract. With this structure of wording, the epistemological framework enables the ethical assessment of cyber risks from shared IoT infrastructure. From the structure of the core, themes and choices/tasks, all participants would be able to assess the risks of individual strategic, functional and operational themes, choices and tasks. This enhances the role of the auditors and the application of due diligence to health policy of IoT systems.

### *3.3 Lifecycle of healthcare systems and how to update task definitions*

In the healthcare systems natural lifecycle, new emerging risks and vulnerabilities would be detected and identified continuously. Such risk and vulnerabilities would require near real-time process for updating the task definitions. IoT systems present technological advantages for such real-time updates, but without a framework for risk assessment of medical systems, such updates could become invisible to the cybersecurity experts. The framework that we presented, enables visibility of such updates for the ethical assessment of cyber risks from shared IoT infrastructure. For example, when one healthcare department changes the 'Strategic theme', the same department would need to provide detailed description of the task/activities that would be performed in the 'Operational choices' further down in the hierarchy of the framework. By applying this framework, all of the departments in healthcare system would be able to continuously review any changes in the activities in the digitalisation of medical systems. This resolves a very important issue in healthcare ethics, because currently, individual departments are responsible for performing specific medical function, but the detailed technological activities of one department are invisible to the remaining departments. The framework we developed resolves this issue and enables all of the departments involved to perform ethical assessment of cyber risks from shared IoT infrastructure and includes visibility of tasks definitions updates in the entire healthcare systems.

## 4 Discussion and main findings

The epistemological framework enables for the ethical assessment to be correlated to the 'themes' and 'categories'. Through the 'themes' and 'categories', the epistemological framework enables the process of identifying ethical conflict of shared risk in the 'categories', 'activities', and 'operational tasks'. Although the integration of IoT systems can be seen as a strategic decision, associated solely to the participant investing in that IoT system, without identifying the shared cyber risks and the operational and digital capabilities for IoT technologies, it would be impossible to verify if the transfer of cyber risk from one healthcare department to another is ethical. For example, consider the impact on Covid-19 vaccine acceptance, if IoT system is hacked, resulting with loss of patients data, and one healthcare department was responsible for installing and maintaining the IoT system that holds shared patients data. The effect of this cyber-attack will cause negative impact the entire healthcare system, especially in terms of using new technologies for shared responsibility. There is also secondary loss of cyber risk that needs to be considered and assessed by all the departments that could be affected. It is impossible to understand the potential secondary loss (e.g., impact on vaccine researchers from vaccine producers brand reputation – i.e., effect on University of Oxford from problems associated with Astra Zeneca, or legal cost of legal proceedings - i.e., EU against Astra Zeneca), without ethical awareness, transparency, and accountability included in the assessment of shared risk, and integrated in the medical and healthcare system design. Therefore, the epistemological framework enables ethical assessment of shared risk, which is crucial for IoT-enabled design.

To design the epistemological framework, we extracted data from the Web of Science Core Collection on the topics of ethics, IoT and cyber risk. Apart from the Web of Science Core

Collection, we also used the Google scholar search engine and other databases of literature, to identify studies that relate the topics of ethics, IoT and cyber risk. We found many studies on different aspects of ethics and IoT, such as health data of children and IoT, or the ethics of cloud computing, even the environmental ethics. However, our detailed search on the Web of Science Core Collection, failed to identify a single study that relates the topic of ethics, IoT and cyber risk to the keywords related to ethics, shared responsibility, health policy and IoT. In our analysis and visualisation of the data records, we also tried searching for related keywords, e.g., searching for 'trust' instead of ethics, or 'privacy' instead of cyber risk. We used statistical methods with R studio to perform bibliometric analysis with the 'bibliometrix' package, and to visualise keywords. Our evaluation is explained in bibliometric analysis chapter and details our findings that the topic of ethical assessment of the shared IoT cyber risks has not been covered extensively.

We followed on this conclusion with a case study and applied epistemological reflexivity to the most prominent models on IoT technologies and healthcare systems. We identified a wealth of knowledge on the assessment of shared risk, related to a diverse set of risks, except the ethical assessment of shared risk emerging from IoT systems. This present a knowledge gap in literature on how digital strategies should be articulated in medical and healthcare systems and how shared risk from connected systems can be assessed, in terms of ethics.

## 5 Conclusion

The findings of this study emerge from the balance between theoretical and technical contributions and are presented in the form of an epistemological framework for ethical assessment of shared risks in complex and coupled IoT systems. The framework enables ethical awareness, transparency, accountability in the assessment of the new emerging IoT cyber risks in healthcare systems. The epistemological framework emerges from a case study research and bibliometric analysis of existing literature. The epistemological framework in this study, utilised and adapted the existing knowledge from keywords emerging from the bibliometric analysis, that included studies conducted prior to the emergence of Covid-19 (and even before the discovery of IoT technologies), and presented a new approach for integrating ethics in the assessment of shared risk in IoT-enabled medical systems. More specifically, epistemological framework enables ethical awareness, transparency, accountability in assessment of shared risks. The epistemological framework is built upon a design that emerges from integrating concepts from models that emerged prior to Covid-19 and IoT, but it incorporates ethical assessment of shared cyber risks that emerge from the IoT technology and Covid-19. This presents a process of shared cyber risks to be assessed through evaluating the IoT operations and differentiates the study from understanding on ethics in using IoT technologies in medical systems, and the associated cyber risks.

### *5.1 Limitations and further research*

Finally, the Covid-19 and analysis of ethics, shared responsibility, health policy and IoT in this article stems from a limited number of existing studies on these topics and there is a disparity with literature on ethics of shared risk from IoT in managing Covid-19, which is still limited. Hence, the balance of comparison on Covid-19 and IoT-enabled medical systems, as subjects in existing literature lack cohesion. Still, the study applied a bibliometric analysis of large research data records to address the ethical problems in shared risks from IoT systems during the Covid-19 pandemic. The framework in this study represents the first attempt that needs to be further developed by researchers in this field in anticipation of a new Disease X event.